\documentclass[prm,twocolumn,floatfix,superscriptaddress,amsmath,citeautoscript,aps,longbibliography]{revtex4-1}
\usepackage{blindtext}
\usepackage{graphicx}
\usepackage{tabularx}
\usepackage{array}
\usepackage{booktabs}
\usepackage{dcolumn}
\usepackage{bm}
\usepackage{color}
\usepackage[normalem]{ulem}
\usepackage{natbib}
\newcommand{\stkout}[1]{\ifmmode\text{\sout{\ensuremath{#1}}}\else\sout{#1}\fi}
\definecolor{magenta}{cmyk}{ 0, 1, 0,0}
\usepackage{amsmath}
\usepackage{float}
\usepackage{bm}
\usepackage{wasysym}
\usepackage{empheq}

\usepackage{hyperref}  
\hypersetup{hypertex=true,
            colorlinks=true,
            linkcolor=blue,
            anchorcolor=blue,
            citecolor=blue}

\newcommand{\NAAG}{NdAuAl$_4$Ge$_2$}

\begin{document}
\title{Magnetic phase diagram and multiple field-induced states in the intermetallic \\triangular-lattice antiferromagnet NdAuAl$_4$Ge$_2$ with Ising-like spins}

\author{Mengru Cong}
\thanks{These authors have contributed equally to this work.}
\affiliation{Shenyang National Laboratory for Materials Science, Institute of Metal Research, Chinese Academy of Sciences, Shenyang 110016, China}

\author{Han Ge}
\thanks{These authors have contributed equally to this work.}
\affiliation{Department of Physics, Southern University of Science and Technology, Shenzhen 518055, China}

\author{Lei Zhang}
\affiliation{Shenyang National Laboratory for Materials Science, Institute of Metal Research, Chinese Academy of Sciences, Shenyang 110016, China}
\affiliation{Department of Physics, Southern University of Science and Technology, Shenzhen 518055, China}
\affiliation{Ganjiang Innovation Academy, Chinese Academy of Sciences, Ganzhou 341000, China}

\author{Weijun Ren}
\thanks{Corresponding author: wjren@imr.ac.cn}
\affiliation{Shenyang National Laboratory for Materials Science, Institute of Metal Research, Chinese Academy of Sciences, Shenyang 110016, China}

\author{Nan Zhao}
\affiliation{Department of Physics, Southern University of Science and Technology, Shenzhen 518055, China}

\author{Tiantian Li}
\affiliation{Department of Physics, Southern University of Science and Technology, Shenzhen 518055, China}

\author{Shanmin Wang}
\affiliation{Department of Physics, Southern University of Science and Technology, Shenzhen 518055, China}

\author{Jinlong Zhu}
\affiliation{Department of Physics, Southern University of Science and Technology, Shenzhen 518055, China}

\author{Jiawei Mei}
\affiliation{Department of Physics, Southern University of Science and Technology, Shenzhen 518055, China}
\affiliation{Shenzhen Institute for Quantum Science and Engineering, Southern University of Science and Technology, Shenzhen 518055, China}
\affiliation{Shenzhen Key Laboratory of Advanced Quantum Functional Materials and Devices,
Southern University of Science and Technology, Shenzhen 518055, China}

\author{Qiang Zhang}
\affiliation{Scattering Division, Oak Ridge National Laboratory, Oak Ridge, Tennessee 37831, USA}

\author{Jieming Sheng}
\affiliation{Department of Physics, Southern University of Science and Technology, Shenzhen 518055, China}
\affiliation{Academy for Advanced Interdisciplinary Studies, Southern University of Science and Technology, Shenzhen 518055, China}

\author{Fei Gao}
\affiliation{Shenyang National Laboratory for Materials Science, Institute of Metal Research, Chinese Academy of Sciences, Shenyang 110016, China}

\author{Bing Li}
\affiliation{Shenyang National Laboratory for Materials Science, Institute of Metal Research, Chinese Academy of Sciences, Shenyang 110016, China}

\author{Zhidong Zhang}
\affiliation{Shenyang National Laboratory for Materials Science, Institute of Metal Research, Chinese Academy of Sciences, Shenyang 110016, China}

\author{Liusuo Wu}
\thanks{Corresponding author: wuls@sustech.edu.cn}
\affiliation{Department of Physics, Southern University of Science and Technology, Shenzhen 518055, China}
\affiliation{Quantum Science Center of Guangdong-Hong Kong-Macao Greater Bay Area (Guangdong), Shenzhen 518045, China}
\affiliation{Shenzhen Key Laboratory of Advanced Quantum Functional Materials and Devices, Southern University of Science and Technology, Shenzhen 518055, China}

\date{\today}

\begin{abstract}
{Geometrical frustration and the enhancement of strong quantum fluctuations in two-dimensional triangular antiferromagnets can lead to various intriguing phenomena. Here, we studied the spin-1/2 triangular lattice antiferromagnet~\NAAG. Thermodynamic and transport properties, such as magnetization and specific heat together with the resistivity measurements were performed. In zero field, two successive phase transitions were observed at $T_{\rm N1}=1.75\pm 0.02$ K and $T_{\rm N2}=0.49\pm 0.02$ K, respectively. Under magnetic field, $\rm XXZ$-type anisotropy was revealed, with the moments pointing along the easy $c$ axis. For $B\parallel c$, multiple field-induced states were observed, and the magnetic phase diagram was established based on the specific heat and magnetization data. The temperature-dependent resistivity measurements indicate that~\NAAG~is a good metal. It is very likely that both the long-range RKKY interactions and the geometrical frustration play an important roles in this case.}

\end{abstract}

\maketitle
\section{Introduction}
Geometrical frustrated quantum magnets have attracted a great amount of attention over the past decades~\cite{FrustratedMag1,FrustratedMag2,FrustratedMag3}. In these frustrated magnets, various exotic magnetic orders with non-trivial quantum spin excitations are expected, especially under external fields. Among different frustrated spin models, the two-dimensional (2D) triangular antiferromagnetic spin lattice is a prototype example. Originally, an entangled quantum spin liquid state that does not exhibit any long-range order, was proposed to be the ground state of this 2D triangular lattice with $S=1/2$~\cite{RVB}. Later numerical studies revealed that the ground state is a 120$^{\rm o}$ magnetic order, where only the nearest neighbor antiferromagnetic interactions were considered~\cite{120-1,120-2}.

Experimentally, 2D spin-1/2 antiferromagnetic triangular lattices have been realized in a variety of systems, including the anisotropic triangular compounds Cs$_2$CuBr$_4$ and Cs$_2$CuCl$_4$ with isosceles triangular layers~\cite{Ono2003,Ono2004,Tsujii2007,Fortune2009,CsCuCl1,CsCuCl2}, and the perovskite materials Ba$_3$CoSb$_2$O$_9$ with ideal equilateral triangular layers~\cite{Shirata2012,Zhou2012,Susuki2013}. Multiple fractional plateau phases, including the $1/3$ and higher fractional states were observed in Cs$_2$CuBr$_4$ and Ba$_3$CoSb$_2$O$_9$ under magnetic fields~\cite{Fortune2009,Susuki2013,BaCoSbO2}. To understand the spin dynamics in these systems, it is important to establish the spin Hamiltonian with determined microscopic spin exchanges parameters. Inelastic neutron scattering in the high field polarized state is always a useful technique, where the quantum normalizations are small. However, for most of these $d$-electron based triangular spin systems, critical fields are very high, and they are usually beyond the limit of most laboratory magnets.

Recently, more attention has been drawn to the 2D geometrically frustrated triangular-lattice with rare-earth ions, such as the oxidized RMgGaO$_{4}$ (R = Yb, Tm) and  the dichalcogenide delafossites NaYbCh$_2$ (Ch = O, S, Se), in which the rare-earth ions form well-separated triangular layers~\cite{NYS,NYSe,YMGO,TMGO}. Intriguing properties such as continuum like fractional spin excitations and  Berezinskii-Kosterlitz-Thouless (BKT) phase transitions were observed in these systems~\cite{TMGO1,TMGO2,TMGO3,NYS2}. In contrast to $d$-electrons, the 4$f$-electrons are more localized, and the exchange couplings are relatively weak. In addition, the large effective $g$-factor arising from the strong spin-orbital coupling, allows a more efficient tuning with external fields. As a consequence of these reasons, the upper critical fields for these rare-earth ion based magnets are typically smaller by almost one order of magnitude, and thus it is much easier to access the whole field temperature phase diagram for the rare-earth based compounds.

In most of these insulating frustrated systems mentioned above, the spin correlations are usually dominated by the nearest-neighbor (NN) spin interactions. On the contrary, metallic compounds can host sizable next nearest-neighbor (NNN) interactions or interactions of even further distance, mediated by the conduction electrons. As these long-range spin interactions are introduced into the triangular spin lattice, more peculiar quantum ground states are expected~\cite{SL1,SL2,SL3,SL4,SL5,SL6}. In addition, the presence of the conduction electrons provides a unique opportunity to explore the novel transport properties related to the possible non-trivial field-induced spin textures, such as the topological Hall effect found in other metallic frustrated lattices~\cite{TopoHall}.

Motivated by this idea, we investigated the rare-earth based intermetallic family RAuAl$_4$Ge$_2$ (R= rare-earth), where the magnetic rare-earth ions are arranged in well-separated triangular lattices~\cite{RAuAl4Ge2}. Previous studies of the isostructure compound CeAuAl$_4$Ge$_2$ indicate that the Ce$^{3+}$ moments are lying in the $ab$-plane, with a ferromagnetic intra-plane interaction, which partly releases the spin frustration~\cite{CeAuAl4Ge2}. By replacing the magnetic Ce ions with Gd and Tb ions, larger magnetic moments and stronger antiferromagnetic interactions are realized~\cite{Gd_TbAuAl4Ge2}. The recent studies about GdAuAl$_4$Ge$_2$ and TbAuAl$_4$Ge$_2$ reveal intriguing phase diagrams, with multiple metamagnetic transitions observed at low temperatures~\cite{Gd_TbAuAl4Ge2}. In addition, specific heat measurements suggest strong magnetic fluctuations persist in the paramagnetic region above the antiferromagnetic phase transition.

In this paper, we present the study of the neodymium (Nd) based triangular-lattice antiferromagnet~\NAAG. We found that the magnetic moments of Nd$^{3+}$ are not constrained in the $ab$-plane. Instead, they point along the $c$-axis with an XXZ-type anisotropy at low temperatures. Statical antiferromagnetic (AFM) orders were established, with two successive transitions at $T_{\rm N1}=1.75\pm 0.02$ K and $T_{\rm N2}=0.49\pm 0.02$ K in zero fields. In addition, multiple magnetic states were found in~\NAAG~with the field applied along the easy $c$-axis.

\section{Experimental method}

\NAAG~single crystals were grown by the metallic self-flux method~\cite{RAuAl4Ge2}. The high purity starting elements neodymium (Nd, Alfa Aesar, 99.99$\%$), gold (Au, Aladdin, 99.99$\%$), aluminium (Al, Alfa Aesar, 99.9$\%$), and germanium (Ge, Aladdin, 99.999$\%$) were mixed together in an alumina crucible with extra Ge and Al as the flux. The mixture was loaded in evacuated quartz tubes and heated up to 1080$^{\circ}\rm C$ followed by a slow cooling procedure to 600$^{\circ}\rm C$. The extra flux were then mechanically removed by the centrifuge at this temperature. Hexagonal-shaped~\NAAG~single crystals of sizes around $2\times2\times0.5~\rm mm^3$ were left inside the growth crucibles. The crystal structure of~\NAAG~was verified by the Bruker APEX-II diffractometer. The magnetic susceptibility and specific heat measurements were performed by the commercial Quantum Design Magnetic Property Measurement System (MPMS) and the Physical Property Measurement System (PPMS) with a $^3$He refrigerator insert.

\section{Results and discussions}
\subsection{Crystal Structure and Crystalline Electrical Field}

\begin{figure}[t!]
  \includegraphics[width=0.4\textwidth]{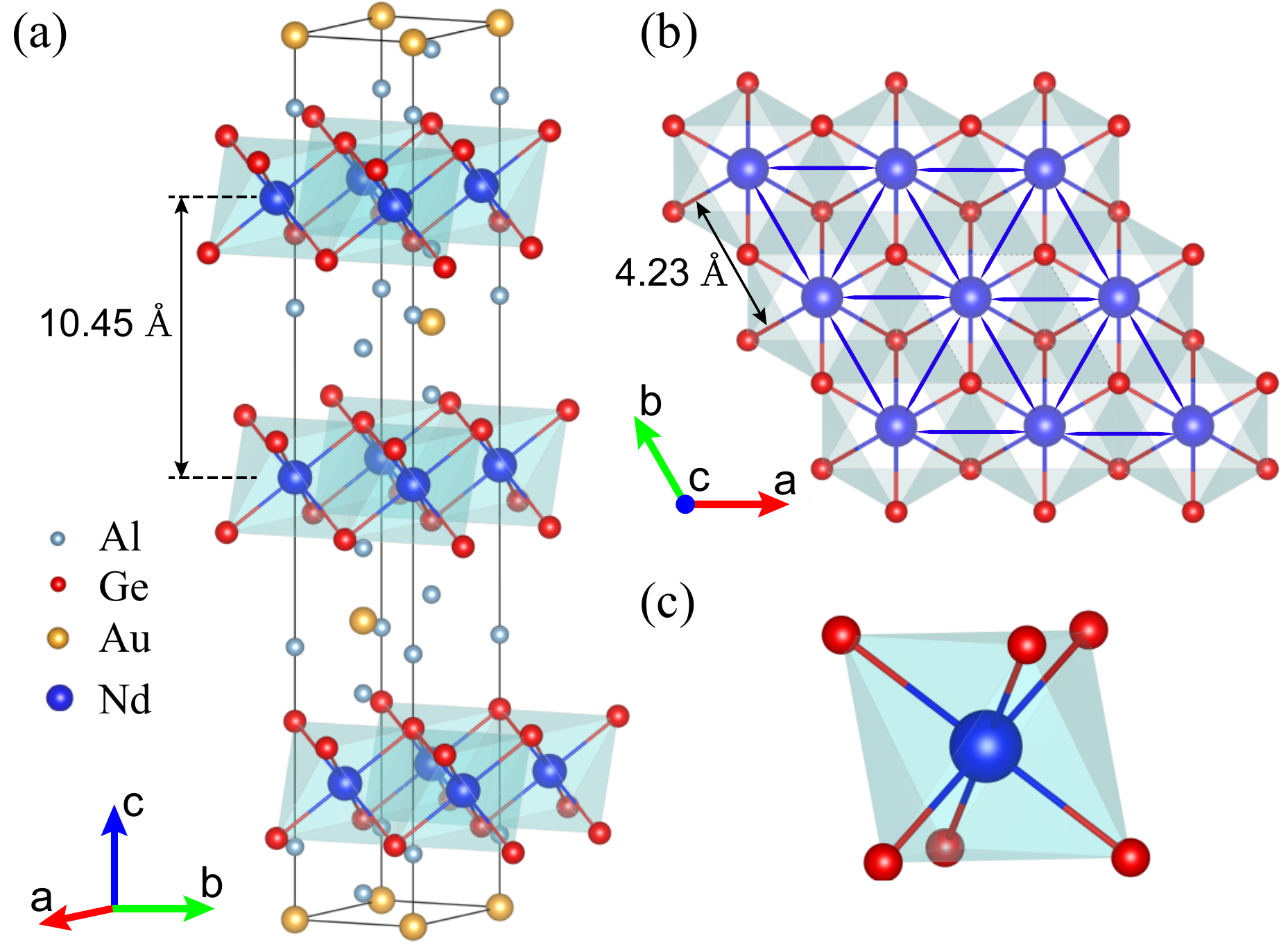}
  \caption{(a) Crystal structure of \NAAG. (b) The triangular layer of magnetic NdGe$_6$ octahedra in the $ab$ plane. (c) NdGe$_6$ octahedra, with the local point symmetry $D_{\rm 3d}(\bar{3}m)$ at the Nd site.}\label{Structure}
\end{figure}

\begin{figure*}[ht!]
 \includegraphics[width=0.9\textwidth]{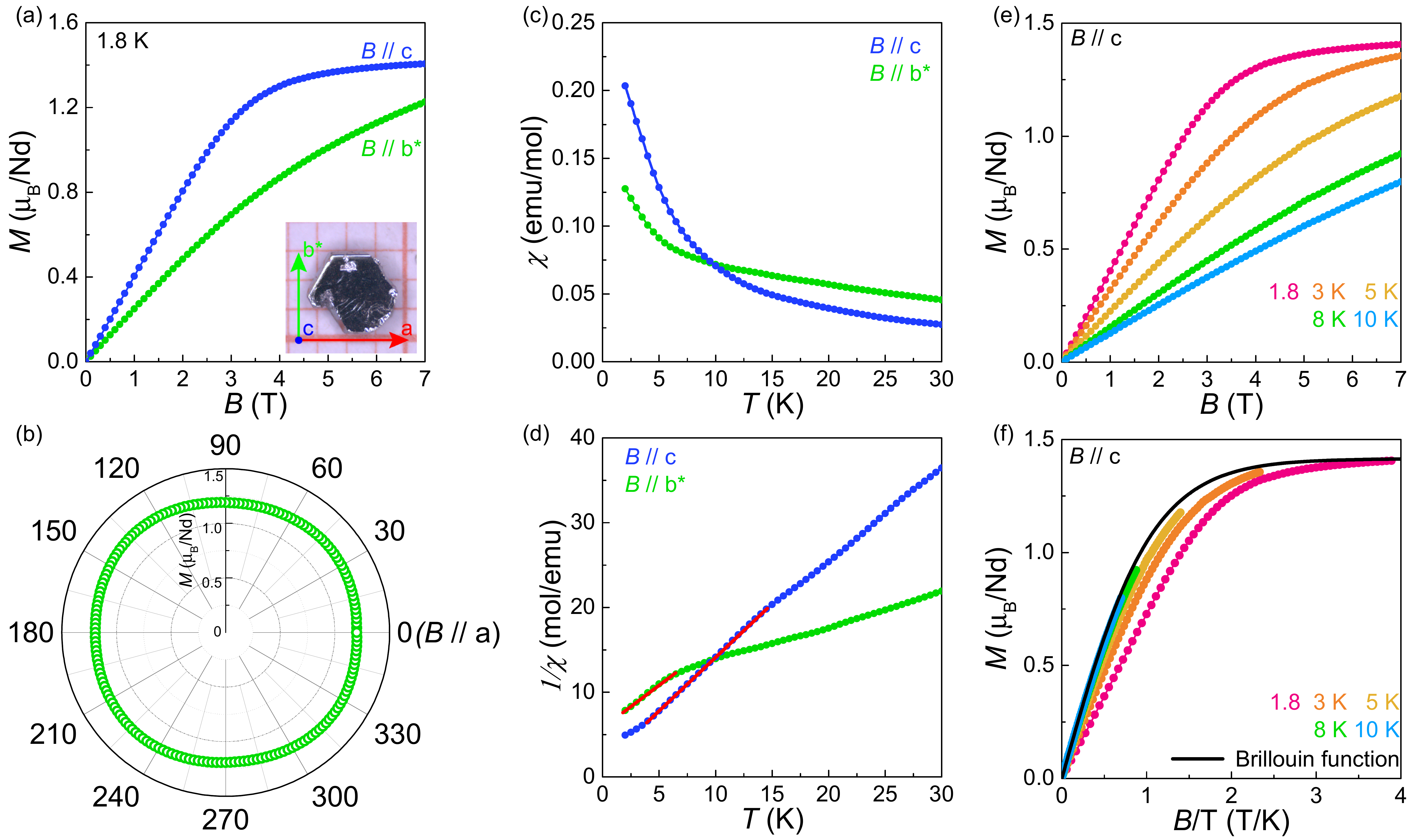}
    \caption{(a) Isothermal magnetization, $M(B)$, measured at 1.8 K with field applied along the $c$ and $b{^*}$ directions. Inset: Single crystals of~\NAAG~grown by the flux method, with sizes about $2\times2\times0.5$ mm$^{3}$. (b) The angle-dependent magnetization measured at $1.8 $ K, $7$ T with the magnetic field applied in the $ab$-plane. (c) Temperature dependence of the magnetic susceptibility $\chi$, (d) and the inverse susceptibility $1/\chi$, measured in field 0.1 T along the $b^*$ and $c$ directions. The red solid lines show the fits base on the Curie-Weiss law. (e) Field dependence of magnetization at different temperatures with field along the $c$-axis. (f) Magnetization measured at different temperatures presented as a function of $B/T$. The black solid line is the calculated Brillouin function corresponding to non-interacting Nd$^{3+}$ ions with $J_{\rm eff} = 1/2$ and the effective g-factor $g_{\rm eff}^{\rm c} = 2.83$ for $B \parallel c$.}\label{MBT}
\end{figure*}

\NAAG~crystallizes in the hexagonal crystal structure (space group: $R\bar{3}\rm m$, No.166) with lattice parameters $a=b=4.2258(4)$ {\AA} and $c=31.359(5)$ \AA~\cite{RAuAl4Ge2}. The crystal structure is shown in Fig.~\ref{Structure}. These magnetic Nd$^{3+}$ ions in the center of a distorted NdGe$_6$ octahedra, form an equilateral triangular network in the crystal $ab$ plane, with the nearest-neighbour distance of about 4.23~\AA~(Fig.~\ref{Structure}(b)). Along the $c$-axis, these triangular planes are alternatively stacked with the ABC type sequence. The non-magnetic Au and Al ions are sandwiched between these triangular planes, resulting in well-separated magnetic layers with an inter-layer distance of about 10.45~\AA~(Fig.~\ref{Structure}(a)).


In~\NAAG, the magnetic Nd$^{3+}$ ions are surrounded by six nearby Ge ions, with a local point symmetry $D_{\rm 3d}(\bar{3}m)$. This local point symmetry lifts the $2J+1=10$ ($L=6, S=3/2$, and $J=9/2$) multiplet states of Nd$^{3+}$ into five doublet crystalline electrical field (CEF) states. As the temperatures drop low enough, the magnetic properties are dominated by the ground doublet states. This enables us to describe the low temperature properties with an effective model of the quantum spin $S=1/2$. In addition, constrained by the local symmetry, these magnetic Nd$^{3+}$ moments are expected to either point along the triple high-symmetric direction (Ising-like along $c$) or lie in the plane perpendicular to the $c$-axis ($\rm XY$-type in the $ab$-plane).

\subsection{Magnetization and Single Ion Anisotropy}

Further investigations of the magnetic ground state are presented in Fig.~\ref{MBT}. As expected from symmetry arguments, these moments are likely to lie either along $c$ or in the $ab$-plane. Thus, two crystal directions ($B\parallel c$ and $B\parallel b^*$, inset of Fig.~\ref{MBT}a) are chosen. Fig.~\ref{MBT}(a) shows the field dependence of the isothermal magnetization, $M(B)$, measured at 1.8 K. For $B\parallel c$, the magnetic moments saturate at higher fields, with $M_{\rm s}\simeq 1.41~\mu_B$/Nd at 1.8 K, 7 T. On the other hand, the magnetic moment keeps increasing until 7 T for $B \parallel b{^*}$. The angle-dependent magnetization with the magnetic field applied in the $ab$-plane is shown in Fig.~\ref{MBT}(b). Nearly isotropic behaviors are found, with the measured moment around 1.23 $\mu_B$/Nd at 7 T for $B \parallel ab$.

The temperature-dependent magnetic susceptibility, $\chi(T)$, with $B \parallel b^*$ and $B \parallel c$ are shown in Fig.~\ref{MBT}(c). For both field directions, no trace of the long-range magnetic order is observed down to 1.8 K. It is interesting to notice that, at higher temperatures, the measured susceptibility for $B \parallel b^*$ is larger than that for $B \parallel c$. However, at temperatures lower than $T_{\rm cross}=8$ K, the $c-$axis turns out to be the easy axis. These anisotropic behaviors are related to the CEF configurations. When the measuring temperatures are lower than the excited CEF states, the magnetic properties are dominated only by the ground doublet states. However, the excited CEF states will contribute more with increasing temperatures. The crossover temperature $T_{\rm cross}=8$ K reflects the energy scale of the first excited CEF level, which is likely to be located around a few millielectronvolts (meV) from the ground doublet state.

The inverse magnetic susceptibility is shown in Fig.~\ref{MBT}(d). The red solid lines are the fit to the Curie-Weiss law, $\chi(T) = {C}/(T-\theta_{\rm CW})$, where $\theta_{\rm CW}$ is the Curie-Weiss temperature, and $C$ is the Curie constant associated with the effective magnetic moment. For temperatures higher than 20 K, the Curie-Weiss fit gives rather large negative values of the Curie-Weiss temperatures. However, these values are not indications of strong antiferromagnetic interactions. Instead, the magnetic susceptibility at these temperatures is greatly affected by the excited CEF levels. Here, we focus on the lowest temperature region, trying to minimize the affect of the CEF splittings. The effective moments and Curie-Weiss temperatures are found out to be $\mu_{\rm eff}^{\rm b^*} = 2.75~\mu_{\rm B}$/Nd, and $\theta_{\rm CW}^{b^*} = -5.38$~K for $B \parallel b^*$, and $\mu_{\rm eff}^{c} = 0.79~\mu_{\rm B}$/Nd, and $\theta_{\rm CW}^{c} = -1.12$~K for $B \parallel c$, respectively. These refined effective moments are quite different from the value $3.62~\mu_{\rm B}$/Nd expected from free Nd$^{3+}$ ions, nor the saturation moments expected from the doublet ground state. This is because that the first excited CEF levels still contribute to these temperatures. However, the negative Curie-Weiss temperatures in both directions indicate antiferromagnetic interactions persist at low temperatures. It need to emphasized again that these values pbtained by fitting the Curie-Weiss law cannot be dealt with quantitatively, but can be used as a useful guide for further neutron scattering experiments.

Isothermal magnetization curves measured at different temperatures are presented in Fig.~\ref{MBT}(e). Although no statical magnetic order has been established down to 1.8 K, indications of antiferromagnetic correlations can be found below 10 K. Shown in Fig.~\ref{MBT}(f) are the measured isothermal magnetization presented as a function of field divided by temperature ($B/T$). Over-plotted is the Brillouin function (black solid line in Fig.~\ref{MBT}(f)) of $S = 1/2$, with $g_{\rm eff}^{\rm c} = 2.83$. We can find that the magnetization measured at 10 K can be described well by the Brillouin function. At lower temperatures, however, a clear deviation is observed, suggesting that the antiferromagnetic interactions begin to play a role.

\subsection{Zero Field Magnetic Transitions}

\begin{figure}[t!]
 \includegraphics[width=0.475\textwidth]{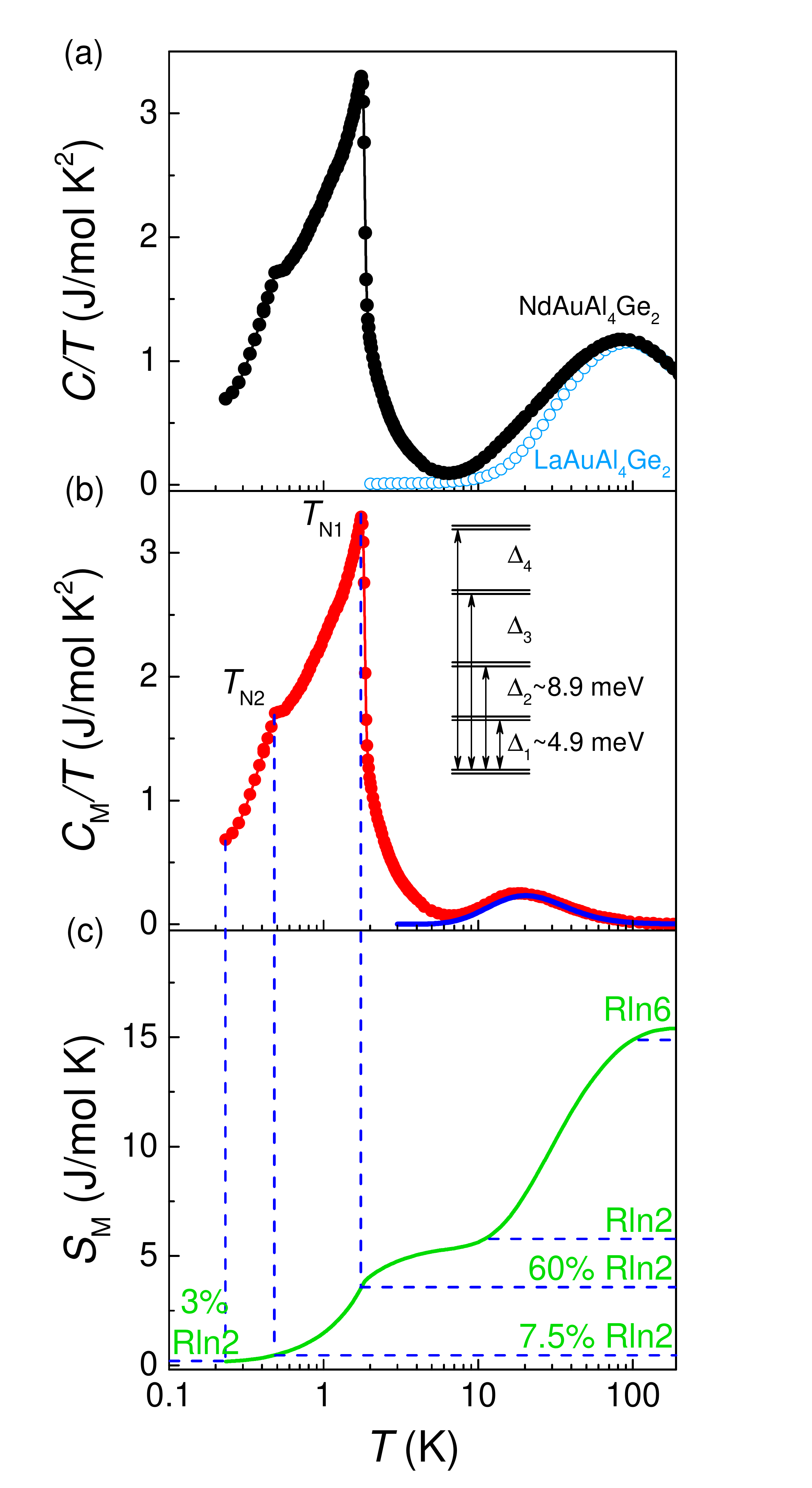}
    \caption{(a) Temperature-dependent specific heat $C/T$ of~\NAAG~(black circles, $C_{\rm Nd}$) and  LaAuAl$_4$Ge$_2$~(blue circles, $C_{\rm La}$) in zero field. (b) Temperature dependence of the magnetic specific heat (red circles, $C_{\rm M}=C_{\rm Nd}-C_{\rm La}$) with the nonmagnetic contributions removed. The blue line is the fit of the Schottky anomaly with the first two CEF excited levels considered.  Inset: A simplified sketch of the CEF configurations, where the first and second excited doublet states are located around  $\Delta_{1}\sim \rm 4.9~meV$ and $\Delta_{2}\sim \rm 8.9~meV$ above the ground doublet state. (c) Temperature-dependent integrated magnetic entropy $S_{\rm M}$ (green line). A full entropy of $R$ln2 and $R$ln6 is released around 10 K and 100 K, respectively.}\label{CT}
\end{figure}

To explore the magnetic properties below 1.8 K, specific heat measurements were performed. The zero-field temperature-dependent specific heat $C(T)$ is presented in Fig.~\ref{CT}(a). The nonmagnetic isostructural compound LaAuAl$_4$Ge$_2$ was synthesized and used to estimate the nonmagnetic contributions (blue circles in Fig.~\ref{CT}(a)). The magnetic specific heat $C_{\rm M}$ with the nonmagnetic contributions subtracted is shown in Fig.~\ref{CT}(b)~(red circles). A sharp peak at $T_{\rm N1}=1.75\pm{0.02}$ K indicates the establishment of the long-range magnetic order. Interestingly, the system goes through another phase transition with a weak shoulder like anomaly appearing at $T_{\rm N2}=0.49\pm{0.02}$ K. The integrated magnetic entropy is presented in Fig.~\ref{CT}(c)~(green line) with 3\% of the zero point entropy $R$ln2 below 0.2 K by linearly extrapolating the specific heat $C/T$ to zero as the temperature approaches zero. A full entropy of $R$ln2 is realized around 10 K, confirming a doublet ground state. However, only about 7.5\% and 60\% of the full entropy are released at $T_{\rm N2}$ and $T_{\rm N1}$, respectively. Similar observations were found in isostructure family compounds CeAuAl$_4$Ge$_2$, GdAuAl$_4$Ge$_2$ and TbAuAl$_4$Ge$_2$, indicating that strong spin fluctuations persist above these phase transitions~\cite{CeAuAl4Ge2,Gd_TbAuAl4Ge2}.

Two scenarios are possible for understanding these successive phase transitions in zero field. One scenario is that the higher temperature transition at $T_{\rm N1}$ is associated with the ordering of the collinear Ising component, while the lower temperature transition at $T_{\rm N2}$ is related to the BKT-type coplanar transition of the XY-components as observations in the perovskites Ca$_3$CoNb$_2$O$_9$ and Ca$_3$NiNb$_2$O$_9$, where the magnetic Co and Ni ions form deformed triangular lattices~\cite{HC1}. It was argued that the easy axis type spatial anisotropy might be responsible for the observation of these two step transitions in zero field~\cite{HC2,HC3,HC4}. An alternative explanation is that only part of the total moments in the triangular layer are ordered at $T_{\rm N1}$, and the intermediate state between $T_{\rm N1}$ and $T_{\rm N2}$ is a partially ordered magnetic state. This partially ordered state was observed in the intermetallic Kagome ice compound HoAgGe with Ising spins lying in the Kagome plane~\cite{HoAgGe}. However, it is difficult to distinguish these two possibilities with specific heat measurements alone, and further neutron diffraction experiments are needed.

For the specific heat in the temperature region between 10 and 100 K, an additional broad peak is observed (Fig.~\ref{CT}(b)). The Schottky like peak is contributed from the excited CEF levels. The analytical expression of the specific heat with two or more excited states is complicated. However, we can start with the partition function,
\begin{equation}\label{Z}
  Z=\sum_{\rm i}g_{\rm i}e^{-\Delta_{\rm i}/k_{\rm B}T},
\end{equation}
where $k_{\rm B}$ is the Boltzmann constant, $g_{\rm i}$ and $\Delta_{\rm i}$ are the degeneracies and energies of different crystal field levels, respectively. Then the internal energy
\begin{equation}\label{Z}
  U(T) =k_{\rm B}T^2\frac{\partial lnZ}{\partial T}.
\end{equation}
From the above equations, the specific heat can be numerically calculated as
\begin{equation}\label{Z}
  C_{\rm Sch}(T) =\frac{\partial U}{\partial T}.
\end{equation}
The experimental data can be well described when the first and the second excited doublet states are taking into consider (the blue line and the inset of Fig.~\ref{CT}(b)). The energy levels of these two excited states are found to be located at $\Delta_1\simeq4.9 \rm~meV$ and $\Delta_2\simeq8.9 \rm~meV$ (inset of Fig.~\ref{CT}(b)). At the meantime, the integrated entropy connected with this Schottky anomaly indicates a full entropy of about $R$ln6 is released at 100 K, in accordance with the expectation that in total three doublet states are involved.

\begin{figure}[t!]
 \includegraphics[width=0.45\textwidth]{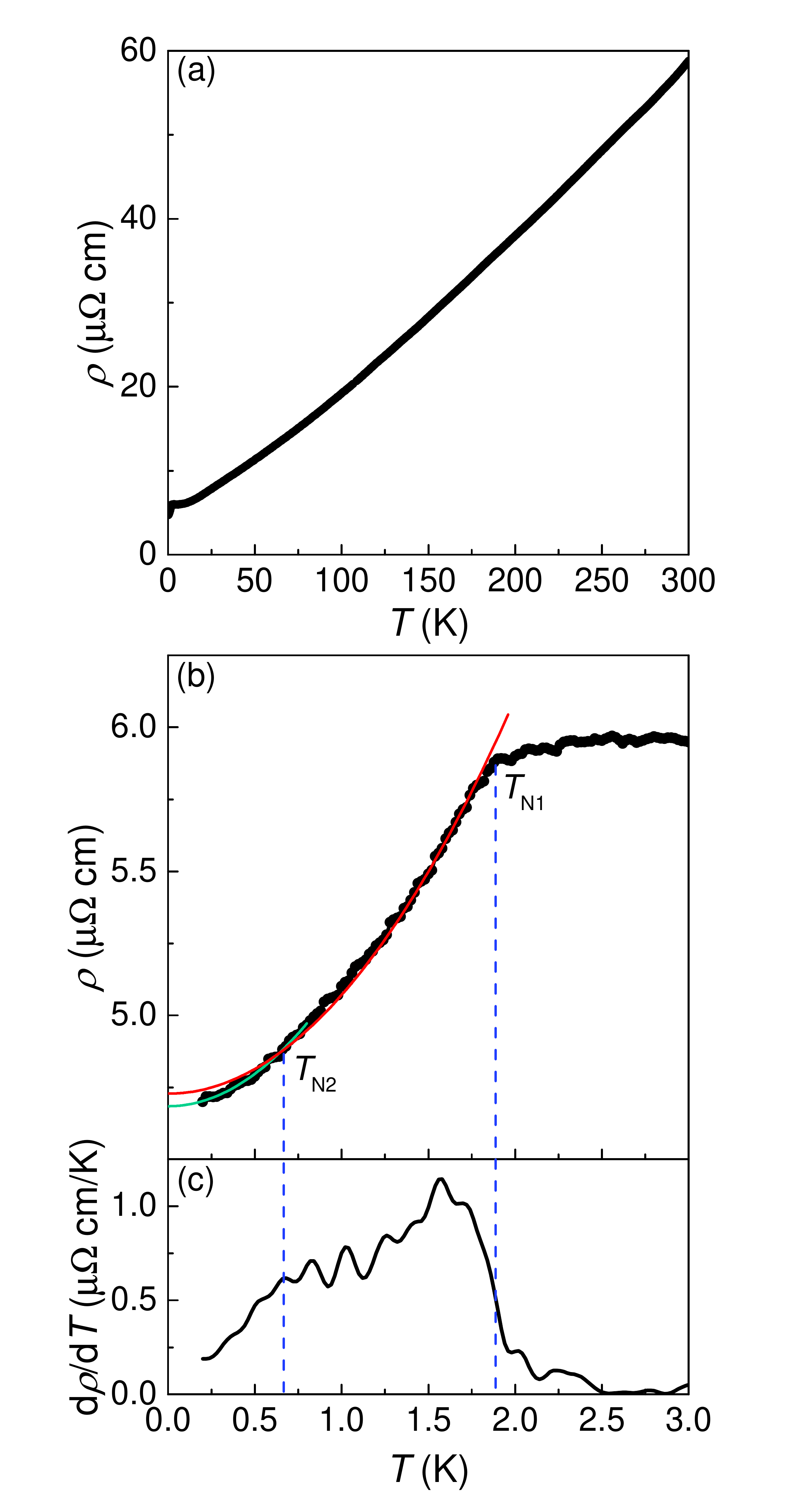}
\caption{(a) Zero field temperature dependence of the electrical resistivity $\rho$ in~\NAAG. (b) Resistivity $\rho (T)$ and (c) temperature derivatives of resistivity ($d\rho/dT$) as a function of temperature at low temperatures region. The red and green lines are fits to the resistivity for Fermi liquid like behaviour in different temperature regions.}\label{RT}
\end{figure}

The temperature-dependent resistivity was also measured, as presented in Fig.~\ref{RT}(a). The resistivity decrease monotonically from about $\rm 60~\mu\Omega\cdot cm$ at 300 K to about $\rm 6~\mu\Omega\cdot cm$ at 3 K, resulting a residual resistivity ratio $RRR=\rho_{\rm 300 K}/\rho_{\rm 3K}\simeq10$. It is worth noticing that no significant upturn was observed down to 2 K, suggesting that the Nd$^{3+}$ moments are well localized and the Kondo effect is very weak in this compound. As the temperature was further decreased, a significant decrease in the resistivity was observed at $T_{\rm N1}$ due to the reduction of spin disorder scattering in the ordered state. On the other hand, no significant anomaly was observed in the resistivity at $T_{\rm N2}$ (Fig.~\ref{RT}(b)). We found that the temperature dependence of the resistivity below $T_{\rm N1}$ can be well described by Fermi liquid-like behavior with $\rho(T)=\rho_0+AT^2$. The second transition at a lower temperature $T_{\rm N2}$ corresponds to a slight change in the coefficient $A$ ($A=0.3425~\mu\Omega\cdot \rm cm/\rm K^{2}$ for  $T_{\rm N1}>T>T_{\rm N2}$, and $A=0.4495~\mu\Omega\cdot \rm cm/\rm K^{2}$ for $T<T_{\rm N2}$) as can be seen from the temperature derivative resistivity of the $d\rho/dT$ (Fig.~\ref{RT}(c)). The residual resistivity $\rho_0$ also reduced from $4.729~\mu\Omega\cdot \rm cm$ to $4.685~\mu\Omega\cdot \rm cm$ as decreasing temperature, indicating a better ordered state below $T_{\rm N2}$.


\subsection{Multiple Magnetic States Induced In Fields}

\begin{figure}[b!]
 \includegraphics[width=0.4\textwidth]{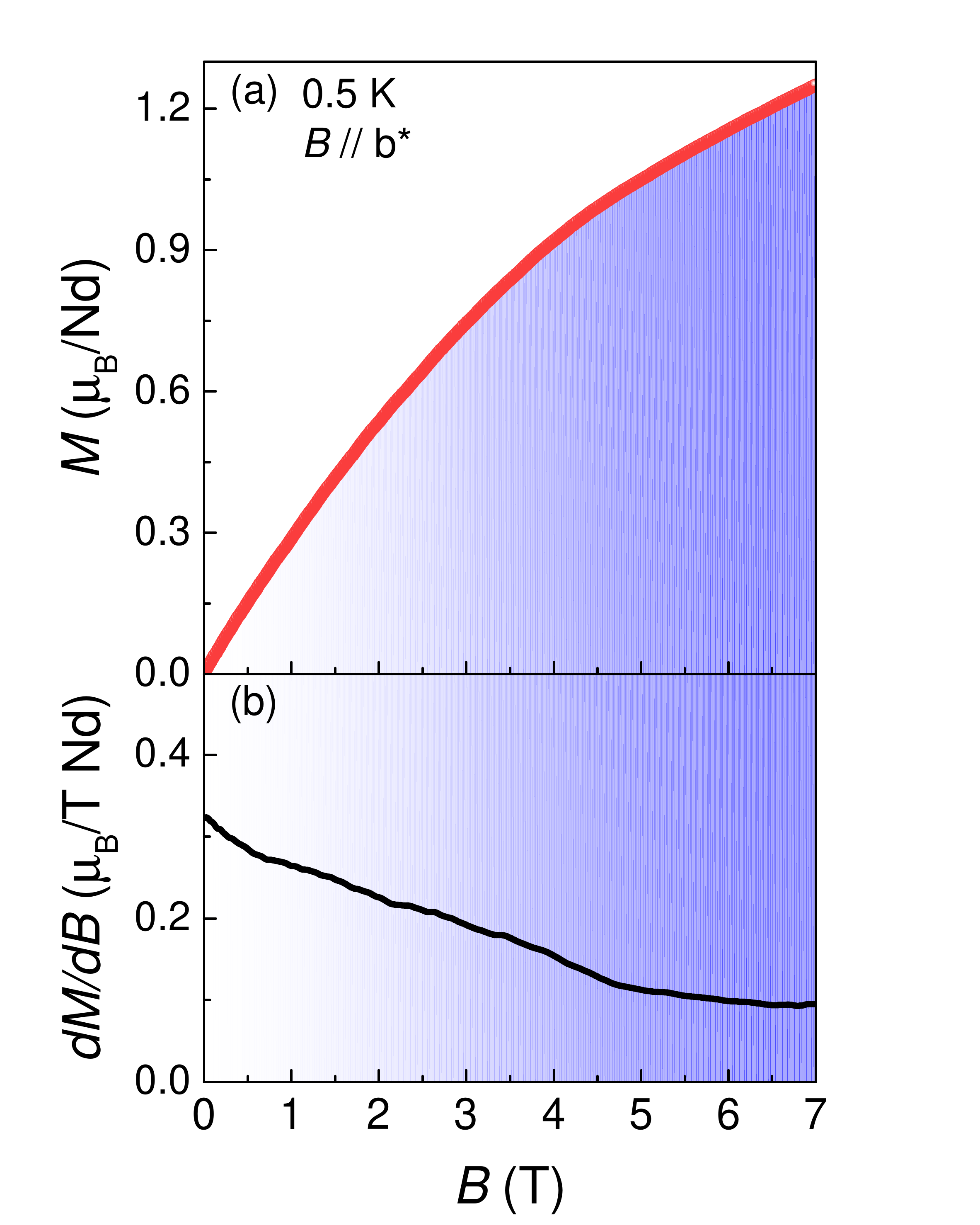}
    \caption{(a) Field-dependent magnetization, (b) and magnetic susceptibility, $dM/dB$, measured at 0.5 K with applied field along the $b^*$ direction.}
    \label{MB_b}
\end{figure}

\begin{figure}[ht!]
 \includegraphics[width=0.46\textwidth]{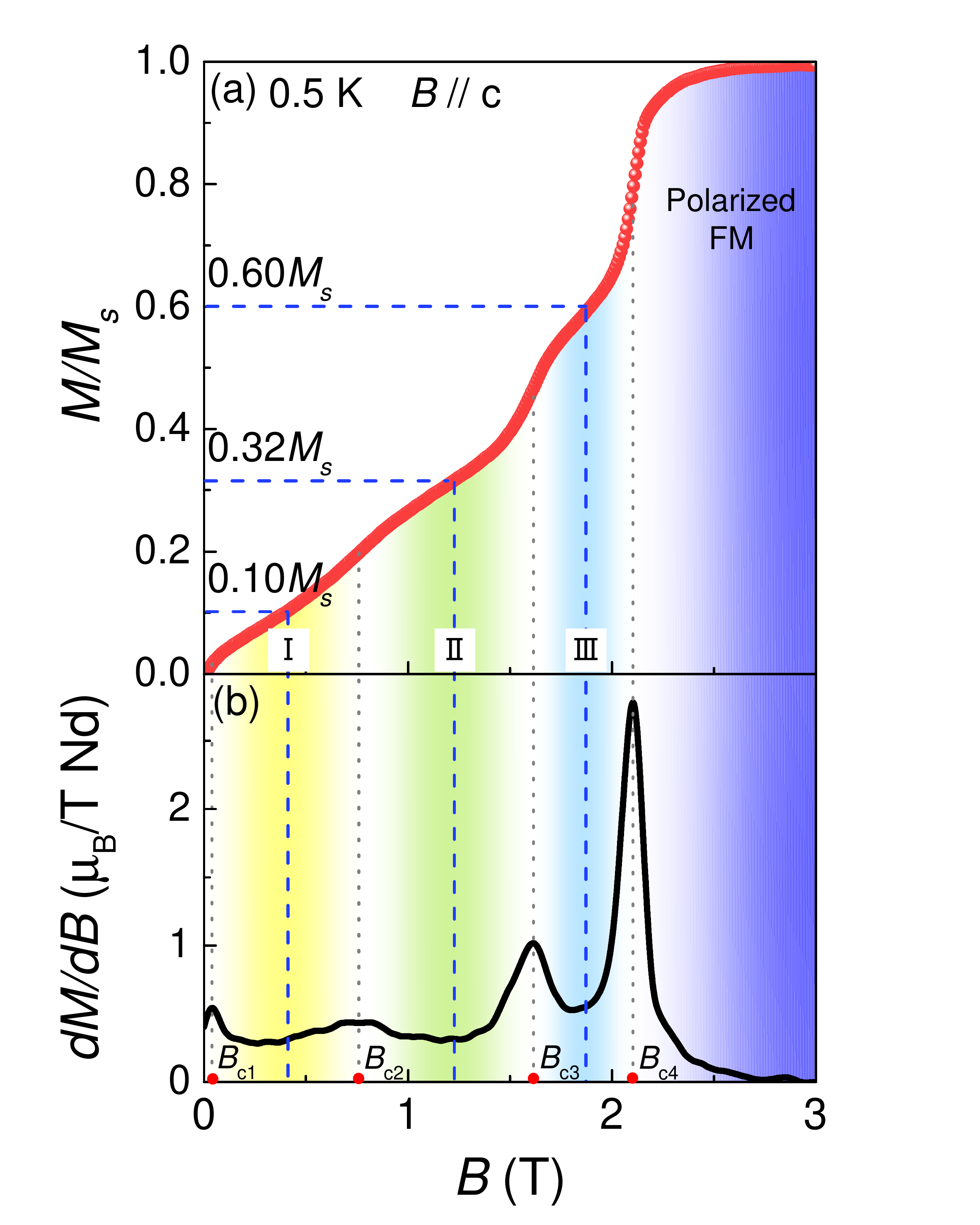}
    \caption{(a) Field-dependent normalized magnetization, $M/M_{s}$, (b) and magnetic susceptibility, $dM/dB$, measured at $0.5$ K with applied field along the c directions. The vertical gray dotted lines indicate the critical fields, and the vertical blue dashed lines are the center between the gray dotted lines.}
    \label{MB}
\end{figure}

\begin{figure}[ht!]
 \includegraphics[width=0.46\textwidth]{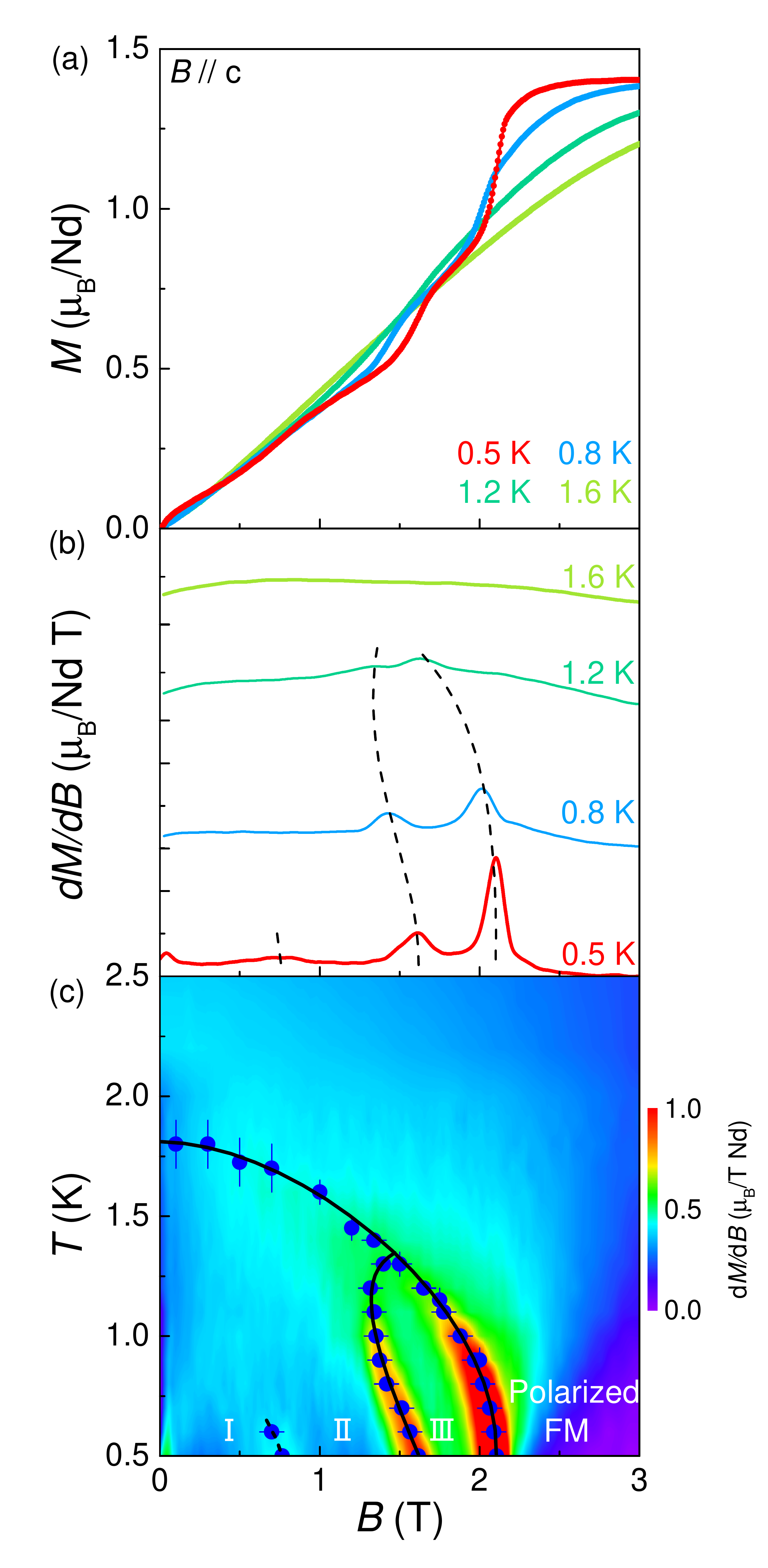}
    \caption{(a) Field dependence of magnetization, (b) and differential magnetic susceptibility, $dM/dB$, at different temperatures below $T_{\rm N}$ with field along the $c$-axis. (c) The field-temperature magnetic phase diagram overlaid on the contour plots of the magnetic susceptibility $dM/dB$ with field along the $c$ direction. The magnetic phase boundaries were extracted through $M(T)$ and $M(B)$ measurements, with the saturation field $B_{s} = 2.1$ T for $B \parallel c$.}
    \label{PD}
\end{figure}

Magnetization measurements below 1.8 K were performed on a home-built Hall sensor magnetometor~\cite{Hallsensor1,Hallsensor2}, integrated with a $^3$He insert. The field-dependent magnetization $M(B)$ for $B\parallel b^*$ at 0.5 K is presented in Fig.~\ref{MB_b}(a). The measured magnetization curve at 0.5 K is qualitatively same as the magnetization measured at 1.8 K shown in Fig.~\ref{MBT}(a). The corresponding differential magnetic susceptibility $dM/dB$ is shown in Fig.~\ref{MB_b}(b), and no obvious anomaly is observed at 0.5 K up to 7 T. This is quite different from the observations in GdAuAl$_4$Ge$_2$ and TbAuAl$_4$Ge$_2$, where easy plane anisotropy was found and multiple magnetic states were induced under external fields applied in the $ab$-plane~\cite{CeAuAl4Ge2,Gd_TbAuAl4Ge2}.

On the contrary, multiple phase transitions are observed for $B \parallel c$, consistent with the expectation that the magnetic Nd$^{3+}$ moments in~\NAAG~lie along the $c$-axis at low temperatures. Shown in Fig.~\ref{MB}(a) are the field-dependent magnetization measured at 0.5 K with $B \parallel c$. These magnetic moments are normalized by the saturation moment $M_{\rm s}=1.41~\mu_{\rm B}\rm /Nd$. It is interesting to find that these multiple phases induced in fields are centered at fractional ratios around $M/M_{\rm s}\simeq0.10$, 0.32, and 0.60, as marked as phase I, II and III in Fig.~\ref{MB}. However, different from the magnetic plateau phases observed in other frustrated triangular systems~\cite{BaCoSbOPRL,NaBaCoPOPNAS}, a finite slope persists in the measured field-dependent magnetization $M(B)$ in these phases. It is also evident that even at the base temperatures, the differential susceptibility $dM/dB$ does not drop to zero in these field-induced phases (Fig.~\ref{MB}(b)). These phase boundaries are defined by the peak like anomalies in $dM/dB$, and the critical fields are found out to be $B_{\rm c1}=0.04$ T, $B_{\rm c2}=0.75$ T, $B_{\rm c3}=1.6$~T, and $B_{\rm s}=2.1$ T.

The distinguished behaviors observed for fields applied along the $c$-axis and the $ab$-plane are consistent with the expectation of previous theoretical reports~\cite{XXZ}, indicating that the magnetic moments are Ising-type, pointing along the $c$-axis. However, we have to mention that~\NAAG~are quite different from most of those previously reported triangular spin lattices such as $\rm Ba_3CoSb_2O_9$~\cite{Susuki2013} and $\rm Na_2BaCo(PO_4)_2$~\cite{NaBaCoPOPNAS}. In those insulating compounds, the nearest-neighbor interactions are usually dominating. Although in~\NAAG, the long-range RKKY interactions mediated by the conduction electrons play a key role, and the second, third neighbor or even longer distance interactions are introduced in the triangular plane. This long-range RKKY interaction in together with the geometrical spin frustration can establish various complex incommensurate (IC) states with sinusoidally modulated magnetic structures in fields~\cite{R2PdSi3}. More interestingly, magnetic lattices with non-trivial spin textures such as skyrmions can be potentially stabilized. However, to verify this scenario, more electrical transport measurements and neutron scattering characterizations under magnetic fields are required in the future.

To clarify the magnetic phase diagram in fields, the magnetization and the differential susceptibility at different temperatures were measured, as presented in Fig.~\ref{PD}(a)-(b). With increasing temperatures, these field-induced magnetic phases gradually melt at higher temperatures. The IC phases I and II at low fields disappear at around 0.6 K, whereas the IC phase III disappears at the temperature around 1.3 K. Finally, smooth field-dependent magnetization was observed above 1.6 K. These phase boundaries extracted based on the peak positions of $dM/dB$ were shown in Fig.~\ref{PD}(b), and the magnetic phase diagram was overlaid with the contour of the susceptibility (Fig.~\ref{PD}(c)).

From the field-temperature phase diagram, we can see that the phase boundary of the IC phase III exhibits an $S$-type feature as the temperature changes (Fig.~\ref{PD}(b) and (c)). The lower critical field of this phase starts from the tri-critical point around 1.46 T at 1.35 K. Then it shifts to the lower field around 1.32 T as the temperature is lowered to 1.15 K. However, as the system temperature is further lowered, the phase boundary shifts to higher fields again, reaching to about 1.62 T at 0.5 K.  Coincidentally, the $S$-type phase boundary has also been observed for the family compound TbAuAl$_4$Ge$_2$~(Fig.8c and Fig.9a in reference~\cite{Gd_TbAuAl4Ge2}). This rare phenomenon is similar to the observations in the solidification of $^3$He from the liquid phase, known as the Pomeranchuk effect~\cite{3He1}, due to the competition between the lattice and magnetic entropy in the superfluid and solid phases. Nevertheless, we also noticed that similar observations were found in the magnetic metallic compound Sr$_3$Ru$_2$O$_7$~\cite{3He2}, where a Fermi-surface distortion was proposed to be responsible for the `spin-dependent Pomeranchuk instability'. For NdAuAl$_4$Ge$_2$, it is still unclear why such a similar phenomenon can be observed. However, it is likely to be raised by the combined effect of geometric spin frustration and the long-range RKKY interactions, and we believe that similar behavior can be expected in other metallic spin frustrated lattices.

\section{Conclusions}

To summarize, single crystals of~\NAAG~with well-separated 2D triangular layers were synthesized. The magnetic properties were investigated through specific heat, magnetization and resistivity measurements. The experimental results, along with the CEF analysis, suggest a doublet ground state with Ising-like moments along the $c$ axis. Long-range magnetic orders are established with successive transitions observed at $T_{\rm N1}=1.75\pm 0.02$ K and $T_{\rm N2}=0.49\pm 0.02$ K in zero field. In addition, multiple incommensurate states were observed, which are likely induced by the long-range RKKY interactions. To resolve the spin configurations in the ground state, as well as these field-induced states, neutron scattering experiments are required. Besides, due to the presence of the RKKY interactions, the couplings between spins are not limited within the nearest neighbors. Both the further neighbor intra-plane interactions and the inter-layer interactions will matter, and it is hard to construct a correct spin Hamiltonian based on the thermodynamics data only. Thus, it is hard to conclude whether these field-induced states are originated from the intra-plane interactions or the inter-layer interactions. To clarify this, inelastic neutron scattering, especially in the high-field polarized state where linear spin wave theory can be applied, is highly demanded.

\begin{acknowledgments}
We Zhentao Wang for useful discussions. The research was supported by the National Key Research and Development Program of China (Grant No.~2021YFA1400400), the National Natural Science Foundation of China (Grants No.~12134020, No.~11974157, No.~52071323, No.~12174175 and No.~12104255), the Guangdong Basic and Applied Basic Research Foundation (Grant No.~2021B1515120015), Shenzhen Key Laboratory of Advanced Quantum Functional Materials and Devices (Grant No.~ZDSYS20190902092905285), Shenzhen Fundamental Research Program (Grant No. JCYJ20220818100405013), and the Shenzhen Science and Technology Program (Grant No.~KQTD20200820113047086).
The Major Science and Technology Infrastructure Project of Material Genome Big-science Facilities Platform supported by Municipal Development and Reform Commission of Shenzhen. Qiang Zhang was supported by the Scientific User Facilities Division, Basic Energy Sciences, US DOE.

\end{acknowledgments}


\end{document}